\documentclass[doublecol]{epl2}
\usepackage{graphicx}
\usepackage{dcolumn}
\usepackage{bm}
\usepackage{amssymb}
\usepackage{amsfonts}
\usepackage{amsmath}

\usepackage{hyperref}

\begin{document}
\title{Recurrent events of synchrony in complex networks of pulse-coupled oscillators}
\author{A. Rothkegel\inst{1,2,3} \and K. Lehnertz\inst{1,2,3}}
\shortauthor{A. Rothkegel \& K. Lehnertz}
\institute{
\inst{1}Department of Epileptology, University of Bonn, Germany\\
\inst{2}Helmholtz Institute for Radiation and Nuclear Physics, University of Bonn, Germany\\
\inst{3}Interdisciplinary Center for Complex Systems, University of Bonn, Germany
}
\pacs{89.75.Kd}{}
\pacs{05.45.Xt}{}
\pacs{84.35.+i}{}
\pacs{87.10.-e}{}

\abstract{
We present and analyze deterministic complex networks of pulse-coupled oscillators that exhibits recurrent events comprised of an increase and a decline in synchrony. Events emerging from the networks may form an oscillatory behavior or may be separated by periods of asynchrony with varying duration. The phenomenon is specific to spatial networks with both short- and long-ranged connections and requires delayed interactions and refractoriness of oscillators.}
\maketitle

Nature possesses various examples of systems composed of many oscillatory elements in which, through mutual interactions, collective dynamics emerges \cite{Winfree1980,Kuramoto1984,Pikovsky_Book2001}. In many biological networks (like the heart or the brain) interactions are short-lasting and may be modeled as pulses, which are fired at a given oscillator phase \cite{Peskin1975,Mirollo1990}. Such pulse-coupled oscillators show, for large arrangements, a variety of collective dynamics with a convergence  of global observables after transients.
Examples range from stable synchronous \cite{Mirollo1990,Timme2002} and asynchronous states \cite{Abbott1993}, phase-locking \cite{Gerstner1996}, partial synchrony \cite{Tsodyks1993,Vreeswijk1996,Kirst2009}, to transitions from asynchronous to synchronous states \cite{Zumdieck2004}. The human brain with its 100 billion neurons, however, may show recurrent changes in global synchrony (as seen, e.g., on the electroencephalogram) associated with physiologic 
or pathophysiologic functioning \cite{Schnitzler2005,Uhlhaas2006}. 

In this Letter, we present a spatial network model of pulse-coupled oscillators (PCOs), which shows collective dynamics with no convergence to either synchrony, partial synchrony, or asynchrony. Instead, the network generates -- even without noise influences or a change of parameters -- recurrent events that are comprised of an increase and decline in synchrony. Events may occur at random looking times and be separated by prolonged periods of asynchrony. Events may also be initiated immediately after completion of the preceding event leading to an oscillatory behavior. The oscillators thus generate a rhythm, which is not directly linked to their intrinsic time-scales (the delay of interactions and the duration of the refractory period) but is an emerging property of the network. 

We study networks of oscillators $n \in N$ with phases $\phi_n(t)  \in [ 0,1 ]$, $\frac{d \phi_n}{d t} = 1$. If for some $t_f$ and some oscillator $n$ the phase reaches 1 ($\phi_n(t_f) = 1$), it is reset to 0 ($\phi_n(t_f^+) = 0$) and we introduce a phase jump in all oscillators $n'$ which are adjacent to $n$ according to some directed graph: 
$$
 	\phi_{n'}(t_f^+) = \phi_{n'}(t_f) +  \Delta\left(\phi_{n'}(t_f)\right).
$$
$\Delta$ denotes the phase response curve. A refractory period of length $\vartheta$ can easily be incorporated into $\Delta$  by setting $\Delta(\phi) = 0$ for $\phi < \vartheta$. 
Consider identical time delays $\tau < \vartheta$ on all outgoing connections of oscillator $n$, such that all its excitations are received when $\phi_n = \tau$. In this case, $n$ can be replaced equivalently with an undelayed oscillator with periodically shifted phase and phase response curve.
We can thus incorporate both refractoriness and delay into an arbitrary phase response curve $\Delta_\epsilon$ (yielding an formally undelayed phase response curve $\Delta$) by setting:
\begin{equation}
 	\Delta(\phi) =\begin{cases} 
			 (1 - \vartheta)\Delta_\epsilon(\frac{\phi - \vartheta + \tau}{1 - \vartheta})& \vartheta - \tau < \phi < 1 - \tau,\\
			0 & \mbox{otherwise}. \\
 	          \end{cases}
\end{equation}
Here $\epsilon$ denotes some coupling strength. Our choice for $\Delta_\epsilon(\phi)$  is the first order in $\epsilon$ approximation of standard integrate-and-fire neurons with excitatory coupling, threshold and eigenfrequency normalized to 1, and exponential charging:
\begin{equation}
 	\Delta_\epsilon(\phi) =\min \left\{ - \epsilon \frac{(1-\alpha)}{\ln(\alpha)} \alpha^{ -\phi} , 1 - \phi \right\}.
\end{equation}
$\alpha$ determines the curvature of $\Delta_\epsilon$ and therefore the amount of leakage. The non-leaky case is obtained for $\alpha \rightarrow 1$. 

In contrast to delayed excitatory couplings without refractory periods \cite{Ernst1995,Ernst1998,Zumdieck2004,Wu2007}
the inclusion of a refractory period allows for synchronous states that are stable even if the sums of incoming connection weights are non-identical. 
To demonstrate this, we consider the dynamical evolution of a perturbation $\{\phi_n(0) | n \in N\}$ with size $\delta_0:= \max\limits_n \phi_n(0) - \min\limits_n \phi_n(0)$. Let $n_>$ ($n_<$) denote an oscillator with maximal (minimal) phase. For small perturbations with $\delta_0 < \vartheta - \tau$, each oscillator will fire and will then stay refractory until every other oscillator has fired once. We can thus define the size of the perturbation $\delta_1 = \max\limits_n \phi_n(t_1) - \min\limits_n \phi_n(t_1)$ for some time $t_1$ at which each oscillator has fired exactly once. $n_>$ will not get excited and will not be overtaken by any other oscillator. As we consider solely advances in phase (no inhibition), we can thus conclude $\delta_1 \leq \delta_0$, or $\delta_{i+1} \leq \delta_{i}$ per induction. 

Moreover, we can give an upper bound for  $\delta_\infty = \lim\limits_{i \rightarrow \infty} \delta_i$.
Since for our choice of the phase response curve, phase-locking of two oscillators is always achieved in a finite number of oscillations, and as we consider only a finite number of oscillators, we can choose a time $t_i'$ after which all excitations are received refractorily. The length of the shortest path between $n_>$ to $n_<$ is bounded by the diameter $d$ of the graph (the longest shortest path). As excitations are received refractorily, the difference in phase for every pair of oscillators along such a path is smaller than or equal $\tau$, and therefore $\delta_\infty = \phi_{n_>} - \phi_{n_<}  \leq d \tau$. In particular, for the case of vanishing time delay, complete synchrony is attracting for all strongly connected graphs (i.e., a graph in which there is a path in each direction for every pair of nodes). For non-vanishing time delay, we have \emph{almost} synchronous states with phases distributed in an interval of size $d \tau$. These states are surrounded by an attracting region in state space (assuming $d \tau < \vartheta - \tau $). 

Whether or not almost synchronous states are reached, when starting from homogeneously distributed phases, depends on the relative size of $\tau$ and $\vartheta$. E.g., for a random network and $\vartheta$ slightly larger than $\tau$ (or for similar PCO networks with refractory periods smaller than the time delay) synchronous states are unstable and we observe asynchronous behavior, while for large $\vartheta$ almost synchronous states are globally attracting. 
 
In the following, we choose $\vartheta$ between these two extremes and investigate the collective dynamics of $\left|N\right|=100.000$ identical PCOs arranged on a spatial network with both short- and long-ranged connections. We connect each oscillator bidirectionally to its $k$ nearest neighbors in a one-dimensional chain with cyclic boundaries. We then remove every directed connection with probability $\rho$ and introduce a directed connection between two randomly chosen, unconnected oscillators, thereby avoiding self-connections (cf. \cite{Watts1998}). To characterize the collective dynamical behavior, we define an order parameter $r(t) = 1/\left|N\right| \left|\sum_{n \in N} e^{2 \pi i \phi_n(t)}\right|$, which yields 1 for complete synchrony and 0 for some balanced distribution of the phases of oscillators.  

\begin{figure}[ht]
\begin{tabular}{c}
 \multicolumn{1}{l}{ \textbf{a}  } \\
 \includegraphics[width = 0.9\columnwidth]{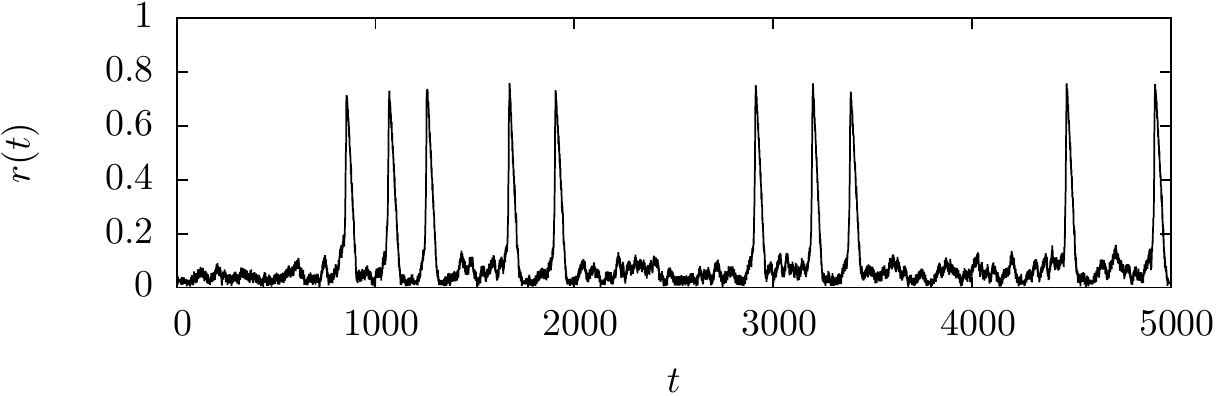} \\
 \multicolumn{1}{l}{\textbf{b}  } \\
 \includegraphics[width = 0.9\columnwidth]{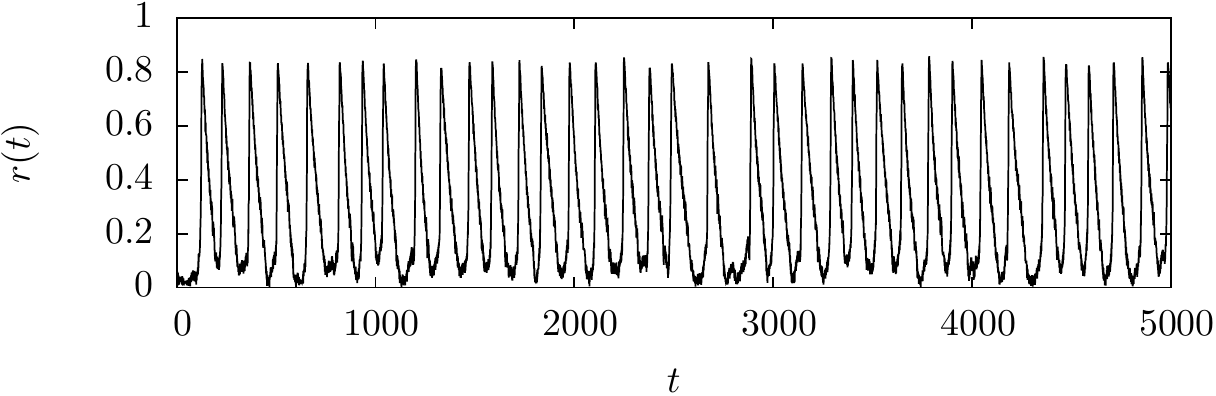} \\   
\end{tabular}
\caption{Examples of recurrent events of synchrony in PCO networks ($\rho = 0.5, k = 50$, $\tau = 0.002, \epsilon = 0.01, \alpha = 0.9$) as assessed by the order parameter $r(t)$. For $\vartheta = 0.027$ ({\bf a}) events are separated by periods of asynchrony with varying duration. For $\vartheta = 0.03$ ({\bf b}) events form an (almost) periodic behavior. Events last for several collective firings of oscillators (for the chosen parameters the number of firings amounts to $\approx 80$).}
\label{fluctuations}
\end{figure}

\begin{figure}[ht]
\includegraphics[width=\columnwidth]{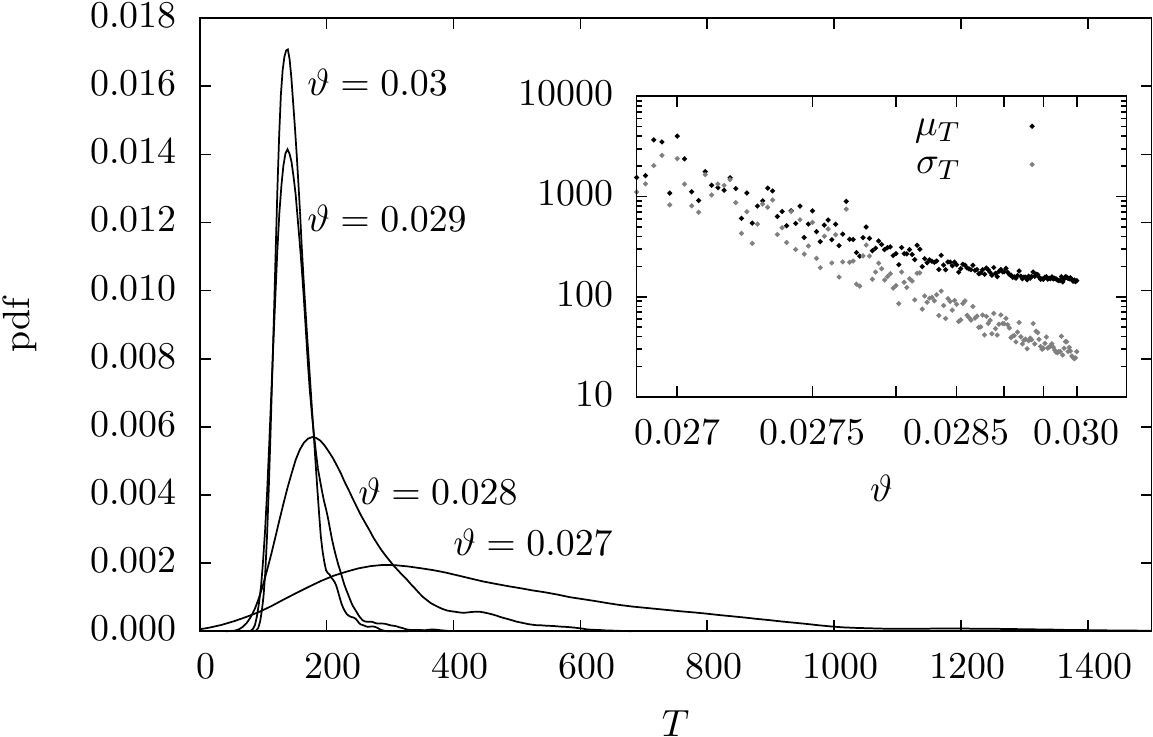}
\caption{Distributions of time periods $T$ between peaks of successive events for different refractory periods $\vartheta$ (observation interval $t \in [0,50.000]$). Inset shows mean ($\mu_T$; black dots) and standard deviation ($\sigma_T$; gray dots) of $T$ in dependence on $\vartheta$.} 
\label{statistics}
\end{figure}

\begin{figure}[ht]
\includegraphics[width = 1.0\columnwidth]{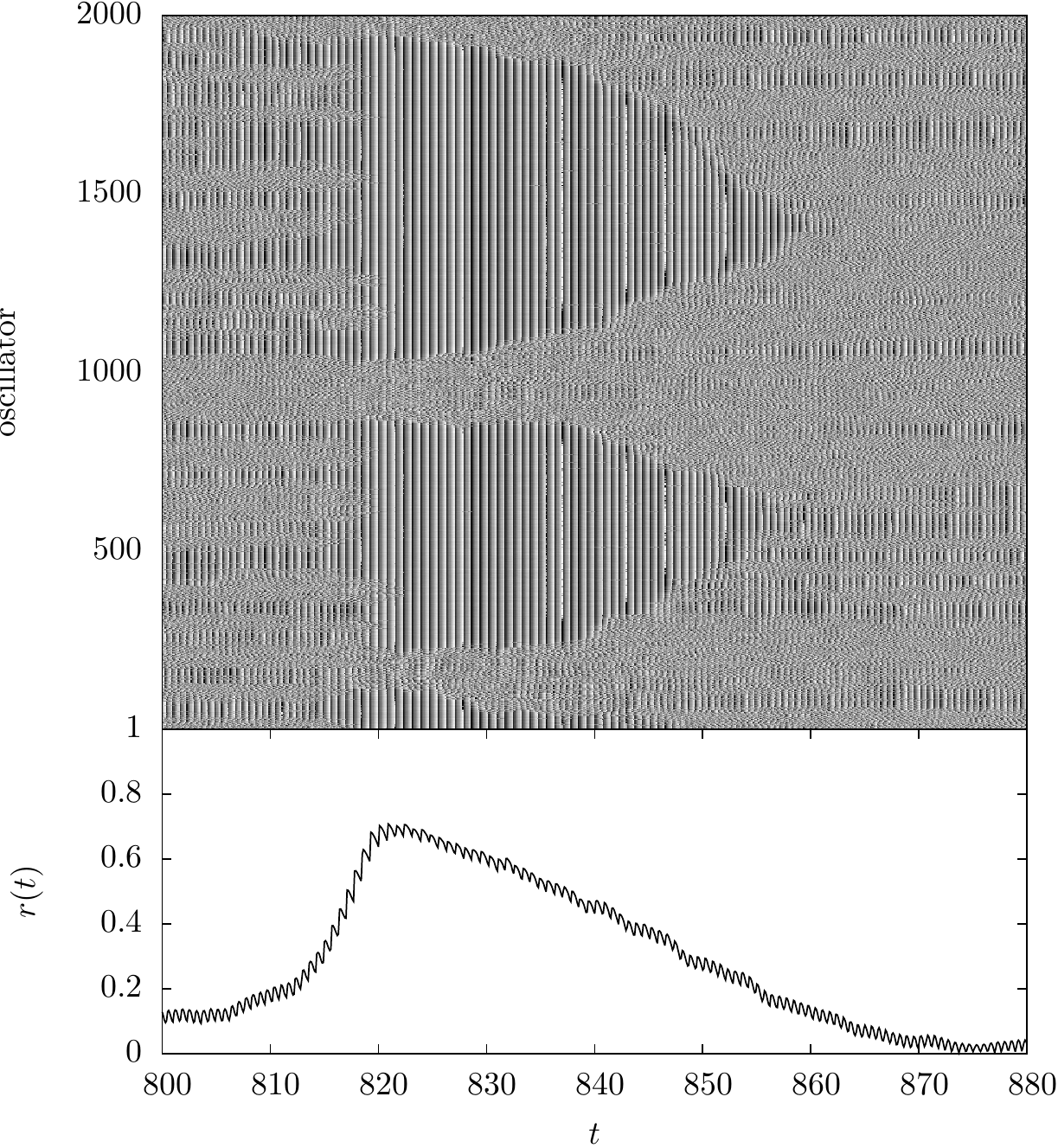}
\caption{Temporal evolution of phases $\phi$ of a part of the network (oscillator 1 to 2.000; $\phi= 0$ black, $\phi = 1$ white) (top) and of the order parameter $r$ calculated for the whole network (100.000 oscillators) (bottom). 
The data are from the first event shown in Fig. \ref{fluctuations}a. The instantaneous frequency $\nu_r$ of the low-amplitude oscillation of $r$ is slightly higher than the eigenfrequency of oscillators and coincides during the event with the frequency $\nu_c$ of collective firings.
\label{spatialDistribution}}
\end{figure}

Starting from homogeneously distributed phases, we observe recurrent events of (partial) synchrony 
in our PCO networks 
(Fig. \ref{fluctuations}). Events are stereotypical and can be characterized by an increase of 
the order parameter 
$r$ up to some peak value ($r<1$) and a subsequent decline. 
For $\vartheta$ a few times larger than $\tau$, events emerge at random looking times, and are separated by prolonged periods of asynchronous firing of oscillators (Fig. \ref{fluctuations}a). For larger $\vartheta$, events emerge almost immediately after the preceding event is completed (Fig. \ref{fluctuations}b). We observe time periods $T$ between peaks of successive events to decrease with increasing refractory period $\vartheta$ (Fig. \ref{statistics}). 
In contrast, the network dynamics during an event and in particular its duration are only marginally affected by the choice of $\vartheta \in [0.027, 0.03]$. In the following, we thus investigate emergence and network dynamics of a single event (Fig. \ref{spatialDistribution}).

In order to characterize the asynchronous behavior prior to the event, we consider the evolution of nearby trajectories in the state space spanned by the phases $\phi_n, n \in N$ using a notion of distance that takes into account their cyclicity: $d(\phi_n,\phi_m)=\min\{| \phi_n - \phi_m| , 1-|\phi_n - \phi_m|\}$ (cf. \cite{Zumdieck2004}). Trajectories are separated whenever an oscillator with phase $\phi$ is excited such that $\Delta'(\phi) > 0$, which is the case for the largest part of our phase response curve (i.e., for $\phi \in  [\vartheta - \tau, 1 - \tau]$). Note that due to the \emph{jump} in the phase response curve at $\phi = \vartheta - \tau$, two nearby trajectories may be separated by an amount which is independent on their distance. For small initial distances, however, this situation becomes increasingly unlikely and we observe an exponential divergence of nearby trajectories, in contrast to inhibitory pulse-coupled networks (investigated in Refs. \cite{Zillmer2006,Jahnke2008}).
Spatial correlations of phases decay over 50 -- 100 oscillators, and phases of oscillators which are linked by a long-ranged connection are uncorrelated. The network dynamics thus resembles spatiotemporal chaos as observed in excitable media \cite{Cross1993}. The chaotic behavior, however, is unstable \cite{Tel2008} and as soon as there is a large enough concentration of phases from distributed oscillators, the network will start to synchronize. However, a stable orbit (as observed e.g. in Ref. \cite{Zumdieck2004}) is not reached in our network. 

\begin{figure}[ht]
\includegraphics[width = 1\columnwidth]{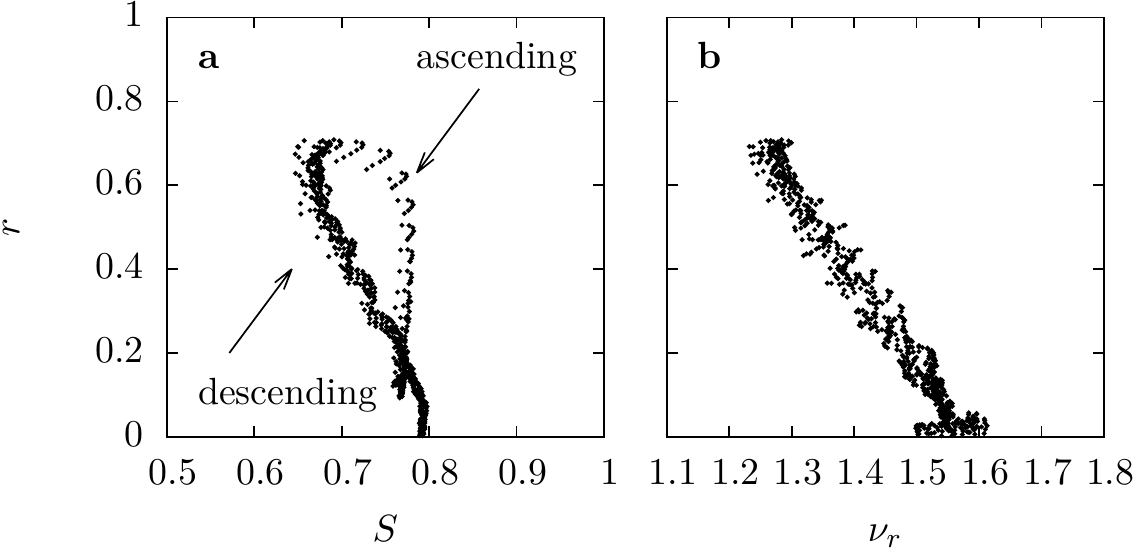}
\caption{Order parameter $r$ versus normalized mean phase distance $S$ ({\bf a}) and instantaneous frequency $\nu_r$ ({\bf b}) for data shown in Fig. \ref{spatialDistribution} bottom. We define $S$ as the mean distance of phases of spatially neighbored oscillators $D_{\rm loc}:=1/\left|N\right| \sum_n d(\phi_n, \phi_{n+1})$ divided by the mean distance between any two oscillators $D:=1/\left|N\right|^2 \sum_{n,m} d( \phi_n ,\phi_m )$. $S$ measures the {\em scatteredness} of the oscillators in the phase cluster. If the oscillators of the phase cluster are homogeneously scattered over the network $D_{\rm loc} \approx D$ and thus $S \approx 1$. If they form connected segments in the network $D_{\rm loc} < D$ and $S$ takes on smaller values. In contrast to $S$, the frequency $\nu_r$ takes on similar values on both the ascending and descending part of the event.}
\label{phasenraum}
\end{figure}

During an event a part of the oscillators forms a single, slightly dispersed phase cluster. The size of the cluster, i. e., the number of contributing oscillators, is reflected in the order parameter $r$ (Fig. \ref{spatialDistribution}).
The chaotic dynamics of the remaining oscillators (asynchronous part) is disturbed by collective firings of the cluster. Oscillators with a phase slightly smaller than the phase of the cluster are more likely to be entrained to the cluster. These oscillators are distributed all over the network (considering the chaotic spatiotemporal dynamics of the asynchronous firing), and their entrainment thus leads to a cluster which is scattered over the network during the ascending part of the event (Fig. \ref{phasenraum}a).
Intuitively, the possibility for entrainment depends on the number of excitations $N_E$ per collective firing received by the asynchronous part and on the frequency $\nu_c$ of collective firings. $N_E$ increases with cluster size but also depends on the spatial distribution of the cluster. On the other hand, $\nu_c$ decreases with cluster size (Fig. \ref{phasenraum}b). This is because excitations via connections which start and end in the cluster are received by refractory oscillators and thus do \emph{not} increase $\nu_c$. Excitations from the asynchronous part to the cluster do increase $\nu_c$ but since their number declines with an increasing cluster size, $\nu_c$ decreases. The impact of the slowing of collective firings on the possibility for entrainment outweighs the impact of the increase of $N_E$. Thus synchronization in our network is not complete. At the peak of the event, the collective firing is too slow to entrain any more oscillators from the asynchronous part despite the large number of excitations per collective firing. The oscillators now form a few, small asynchronous segments which are embedded into synchronous segments (belonging to the phase cluster). 

During the descending part of the event, asynchronous segments grow (Fig. \ref{spatialDistribution}). 
Consider neighboring oscillators well inside a synchronous segment. These oscillators receive excitations via long-ranged connections from oscillators which are distributed homogeneously over the asynchronous part. The resulting phase shifts thus differ only marginally, and these oscillators remain in the cluster.
In contrast, oscillators located near the boundary of a synchronous segment receive (in addition to the excitations via long-ranged connections) correlated excitations via short-ranged connections from the asynchronous part. The resulting phase shift decreases with the distance to the boundary, and these oscillators are torn out of the cluster, leading to a growing of asynchronous segments. The cluster is thus composed of a few connected segments, in contrast to the scattered spatial distribution during the ascending part (Fig. \ref{phasenraum}a). Note that the growing of asynchronous segments can only be observed for networks for which in the initial lattice nodes are connected to at least $k=10$ nearby nodes. For nearest-neighbor coupling, asynchrony spreads via both short- and long-ranged connections which does not lead to events of synchrony.

With a decreasing cluster size, the frequency of collective firings $\nu_c$ increases again (Fig. \ref{phasenraum}b). Nevertheless, entrainment to the cluster is not possible. This is because $N_E$ depends on the spatial distribution of the cluster. For a given cluster size, $N_E$ is large if the oscillators of the cluster are scattered as here excitations via both short- and long-ranged connections contribute. $N_E$ is small if the oscillators of the cluster forms connected segments as here mainly excitations via long-ranged connections contribute. Therefore the cluster is not able to entrain oscillators from the asynchronous part as it was the case during the ascending part of the event. The network thus returns to asynchronous firing, concluding the event.

Spatial networks of spiking elements can exhibit a variety of dynamical behaviors \cite{Hopfield1995,Coombes2005, Jirsa2009,Oestborn2009,Rothkegel2009}. We here reported on a novel phenomenon, namely recurrent events of synchrony, emerging robustly from networks of identical, deterministic oscillators. 
The phenomenon relies on both refractoriness in the local dynamics and delayed excitatory coupling and is specific to networks with both short- and long-ranged connections. With adjusted parameters of oscillators it can be observed for replacement probabilities $ 0.35 \lesssim \rho \lesssim 0.60$. Furthermore, recurrent events of synchrony can be observed for two-dimensional spatial PCO networks and for similar phase response functions without a jump. Nevertheless, the precise dynamical origin of the events needs further investigation.

Complex networks with short- and long-ranged connections have frequently been used to model brain functions and dysfunctions (see, e.g., \cite {Lago2000,Roxin2004,Netoff2004,Percha2005,Bogaard2009,Stam2010}) since these networks represent a simple approximation of the synaptic wiring in the brain \cite{Sporns2004b,Bassett2006b}. In our network, events of synchrony may emerge at random looking times, with prolonged inter-event periods of asynchronous firing of oscillators. This collective phenomenon resembles the dynamics of epileptic brains with seizures as recurrent rare events of overly synchronous neuronal activity. In contrast to previous modeling approaches to the dynamics into and out of seizures that rely either on changes in some control parameter or on noise to generate transitions \cite{Soltesz2008}, our network is capable to transit spontaneously despite the deterministic setup. Our approach may thus significantly improve recent modeling approaches of the disease \cite{Lytton2008}.
Events of synchrony in our network may also occur (almost) periodically, where the oscillators generate a rhythm, which is not directly linked to their intrinsic time-scales but is an emerging property of the network.  The described mechanism that leads to the oscillation may advance understanding of the generation of the respiratory rhythm, which persists even in small slices of mammalian brains after the attenuation of postsynaptic inhibition \cite{Feldman2006}.
\acknowledgments

We are grateful to Stephan Bialonski and Ulrike Feudel for helpful comments. This work was supported by the Deutsche Forschungsgemeinschaft (LE 660/4-1).

\end{document}